\documentclass[journal]{IEEEtran}
\usepackage{graphicx}
\usepackage{subfigure}  
\usepackage[numbers,sort&compress]{natbib}
\usepackage{url}
\usepackage{amsmath}
\usepackage{enumerate}
\usepackage{siunitx}
\usepackage{breakurl}
\usepackage{epstopdf}
\usepackage{pbox}
\usepackage{booktabs}
\usepackage{threeparttable}
\usepackage{textcomp}
\ifCLASSINFOpdf

\else

\fi
\begin{document}

\title{An Energy Efficient Authentication Scheme using Chebyshev Chaotic Map for Smart Grid Environment}

\author{Liping Zhang,
        Yue Zhu,
        Wei Ren*, ~\IEEEmembership{Member,~IEEE,}
        Yinghan Wang,
        and Neal N. Xiong, ~\IEEEmembership{Senior Member,~IEEE}

\thanks{Liping Zhang, Yue Zhu, Yinghan are with the School of Computer Science, China
University of Geosciences, Wuhan, China 430074.}
\thanks{Wei Ren is with the School of Computer Science, China University of Geosciences, Wuhan, China;
Guangxi Key Laboratory of Cryptography and Information Security,
Guilin, P.R. China 541004; and Key Laboratory of Network Assessment
Technology, CAS, (Institute of Information Engineering,
Chinese Academy of Sciences, Beijing, P.R. China 100093); e-mail: weirencs@cug.edu.cn.}
\thanks{Neal N. Xiong is with Department of Mathematics and Computer Science, Northeastern State University, OK 74464, USA}
\thanks{Manuscript received August 20, 2020; revised ...}}

\maketitle

\begin{abstract}
Smart grid (SG) is an automatic electric power transmission network
with bidirectional flows of both energy and information. As one of
the important applications of smart grid, charging between electric
vehicles has attracted much attentions. However, authentication
between vehicle users and an aggregator may be vulnerable to various
attacks due to the usage of wireless communications. Although
several authentication schemes are proposed for smart grid
environments with privacy protection, some of them still impose
security issues such as anonymity absence. In addition, most of the
existing authentication schemes have not thoroughly tackled the
charging peak in their design, so these schemes cannot guarantee the
requirement in terms of low energy consumption in smart grid
environments. In order to reduce the computational costs yet
preserve required security, the Chevyshev chaotic map based
authentication schemes are proposed. However, the security
requirements of Chebyshev polynomials bring a new challenge to the
design of authentication schemes based on Chebyshev chaotic maps. To
solve this issue, we propose a practical Chebyshev polynomial
algorithm by using a binary exponentiation algorithm based on square
matrix to achieve secure and efficient Chebyshev polynomial
computation. We further apply the proposed algorithm to construct an
energy-efficient authentication and key agreement scheme for smart
grid environments. Compared with state-of-the-art schemes, the
proposed authentication scheme effectively reduces the computational
costs and communication costs by adopting the proposed Chebyshev
polynomial algorithm. Furthermore, the ProVerif tool is employed to
analyze the security of the proposed authentication scheme. Our
experimental results justified that our proposed authentication
scheme can outperform state-of-the-art schemes in terms of the
computational overhead while achieving privacy protection.
\end{abstract}


\IEEEpeerreviewmaketitle

\section{Introduction}

\IEEEPARstart{S}{mart} grids create a widely distributed network of
automated energy delivery by using two-way flows of electricity and
information \cite{1}. Compared with the traditional grid
infrastructure, smart grids are more efficient, secure and reliable.
As one of the important applications of smart grids, the
vehicle-to-grid (V2G) network has attracted more and more attentions
due to the electric vehicles that are promising in reducing
pollution and integrating renewable resources \cite{2019Vehicle}. As
shown in Figure 1, the V2G network contains three entities - power
grid, aggregator (AGT), and electric vehicles (EVs). The power grid
generates electricity from new renewable sources, such as solar and
wind, and then sends the resulting electricity to the charging
stations.

\begin{figure}[!t]
    \flushleft
    \setlength{\abovecaptionskip}{-0.01cm}
    \includegraphics{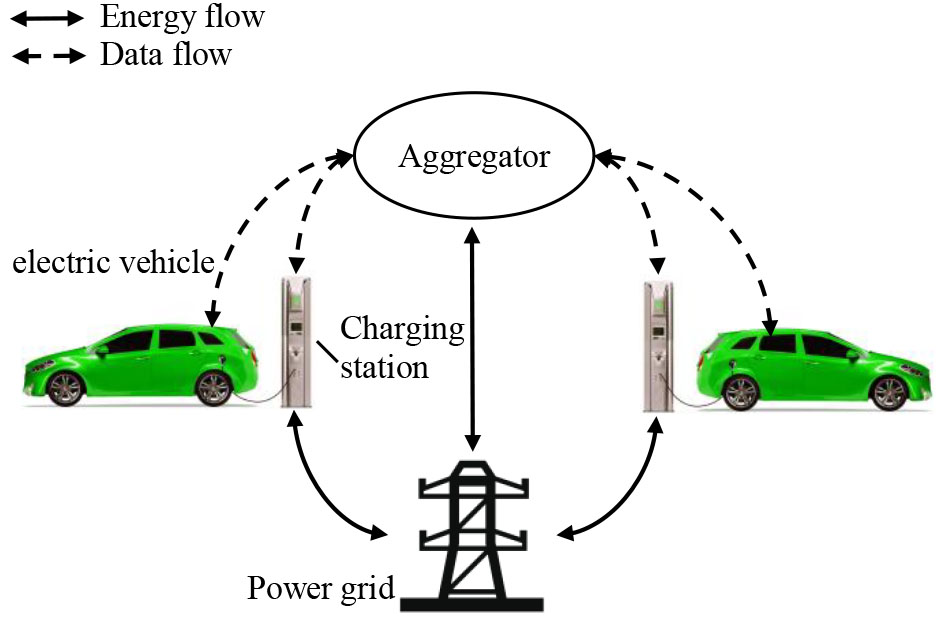}
    \caption{the System Model of the Vehicle-to-Grid Network.}
    \label{fig1}
    \vspace{-0.5cm}
\end{figure}

As an intermediary between the power grid and the EVs, the
aggregator monitors and collects the current state of the EVs,
optimizes and adjusts the EV charging plan, and minimizes the
charging energy cost of electric vehicles. When an electric vehicle
owner (EVO) intends to charge vehicle battery, he/she first logs in
to the system using a smartcard and credentials through EV onboard
unit \cite{0Provably}. Then the EV will send a request to the AGT to
establish communication between them. After that, EVO can complete
the payment by interacting with the charging station.

However, the wireless communication technology adopted in V2G
networks may cause various security threats \cite{4Adaptive,Lin2006An}. User private and
sensitive information (e.g. the user identity, location and route of
the vehicle) may be compromised by several attacks, such as
impersonation attacks, message modification attacks, and etc. In
practice, an attacker may utilize public channels to invade user
privacy information and impersonate users to obtain services through
V2G networks. Furthermore, a malicious station may impersonate
others and charge more from users. Therefore, a security mechanism
should be provided to protect the user privacy information during
the charging process in smart grid environments.

Several authentication schemes have been explored to achieve secure communication \cite{0Provably,8957622,62017A,2010Analyzing,8938138,2010The,92020A}. In order to
provide high security, public-key cryptography has been applied in
the design of the authentication scheme for secure vehicle charging services in smart grid environments. Despite the
fact that the public-key cryptography solutions enhance the
security, these existing authentication schemes still face another
challenge, i.e., energy limitation. The EV charging increases the
electric loads, potentially amplifying peak demand or creating new
peaks in electricity demand \cite{2010Analyzing,2010The}. One
possible scenario is that many EV owners go to charging stations
after work to recharge their EVs, creating a charging peak in a
short period of time. These charging activities can seriously affect
smart grid systems, which causing network losses and increasing the
likelihood of blackouts \cite{2011Real}. Furthermore, when the
charging requirements received by charging stations increase
significantly, it may cause congestion and then greatly affects the
convenience of users. However, most of the existing authentication
schemes do not fully consider the charging peak in their design and
fail to achieve a delicate balance between security and performance
due to the usage of time-consuming operations such as expensive
bilinear operations. Therefore, how to design an energy efficient
authentication scheme for smart grid environments remains a
challenging work.

\vspace{-0.4cm}
\subsection{Motivation}
Although several authentication schemes are proposed for smart grid
environments, some of them still impose security issues such as
anonymity and perfect forward secrecy. In addition, to avoid the congestion caused by charging peak and provide more security and convenient services to EVOs, the authentication scheme should be lightweight. However, most existing authentication schemes have to rely on
time-consuming operations to ensure security which may not be
suitable for smart grid environments.

Chebyshev polynomials cryptosystem, as a promising method, is
expected to reduce computational costs while preserving high
security. However, the public key algorithm based on Chebyshev
polynomial to deal with real numbers is not secure \cite{1487666}.
To solve this problem, the Kocarev \textit{et al}.
\cite{2005Public0} extended Chebyshev polynomials from real fields
to finite fields and finite rings, making the public key algorithm
more secure and practical. Later, Chen \textit{et al}.
\cite{2011Security12} proved that the Chebyshev polynomials
$T_{n}(x)$ mod \textit{N} is safe when modulus \textit{N} is a
strong prime number satisfying \textit{N}-1=2$p_1$ and
\textit{N}+1=2$p_2$ , where $p_1$ and $p_2$ are also prime numbers.
Therefore, modulus \textit{N} should be carefully selected to ensure
that Chebyshev polynomials can produce sequences of sufficient
period to resist violent attacks. However, in the existing Chebyshev
polynomials algorithms such as the algorithms adopted in
\cite{200712A,201012An}, the parameter \textit{n} of Chebyshev
polynomials $T_{n}(x)$ mod \textit{N} is constructed with small
primes which brings a new security challenge to the design of
authentication schemes based on Chebyshev chaotic maps.

These issues inspire us to design a practical Chebyshev polynomials
algorithm and then applied it to construct an energy efficient
authentication scheme for smart grid environments.

\vspace{-0.4cm}
\subsection{Contribution}
In this study, we focus on solving the energy limitation issue with
privacy protection during authentication and key negotiation process
in smart grid environments. The main contributions of our work are summarized as follows:

1) To solve the security challenge of Chebyshev polynomials
algorithm in authentication scheme design, we propose a practical
Chebyshev polynomial algorithm that adopts a binary exponentiation
algorithm based on square matrix to achieve secure and efficient
Chebyshev polynomial computation. The proposed algorithm guarantees
the security requirements that are proved by Chen \textit{et al}
\cite{2011Security12}. Also, our experimental results also justified
that the proposed algorithm is an efficient Chebyshev polynomial
algorithm, thus it can be applied in our designated authentication
scheme to achieve both security and efficiency.

2) Based on the proposed Chebyshev polynomial algorithm, we further
construct an energy efficient authentication scheme for smart grid
environments. The proposed scheme achieves fast mutual
authentication and key agreement with anonymity. Furthermore, we
employ an automatic verifier named ProVerif to analyze the security
of the proposed scheme. The security analysis demonstrates that our
proposed authentication scheme can resist known attacks.

3) The proposed authentication scheme is a lightweight
authentication scheme since only the efficient Chebyshev polynomials
and hash functions are adopted during the authentication and key
negotiations process. The performance analysis shows that the
proposed authentication scheme is more efficient in comparison with
the state-of-art schemes.

\vspace{-0.4cm}
\subsection{Organization}

The rest of this paper is organized as follows. Section II describes
the related work. The mathematical background of Chebyshev
polynomial is described in Section III. In Section IV our proposed
Chebyshev polynomial algorithm is introduced in detail. The proposed
scheme is presented in detail in Section V. In Section VI, the
security of the proposed scheme is analyzed. The performance of the
proposed scheme is discussed In Section VII. And the paper is
concluded in Section VIII.

\section{Related Work}
In recent years, various authentication and key agreement schemes
using public-key cryptography are proposed for smart grids. To
protect the private data transmitted in the smart grid, Wu and Zhou
\cite{2011Fault} proposed a key management scheme using public-key
cryptosystem and symmetric key cryptosystem. However, Xia and Wang
\cite{6205351} demonstrated that their scheme \cite{2011Fault} was
vulnerable to man-in-the-middle attacks. Although another key
distribution scheme was proposed by Xia and Wang \cite{6205351} to
overcome the weakness of \cite{2011Fault}, the new scheme
\cite{6205351} suffered from impersonation attacks
\cite{Park2013Security}. Later, Tsai and Lo \cite{2016Secure}
presented an anonymous key distribution scheme using identity-based
signature. However, their scheme \cite{2016Secure} failed to resist
privileged-insider attacks \cite{2017Secure}.

To further enhance the security and reduce the computational costs,
elliptic curve cryptography (ECC) is adopted in the design of
authentication schemes. Based on ECC, He \textit{et al}.
\cite{2016Lightweight} proposed a key distribution scheme for smart
grids which achieved users anonymously and reduced the computational
costs. Odelu \textit{et al}. \cite{7549086} also presented a key
agreement scheme using ECC and analyzed their scheme with the
CK-adversary model. But their scheme \cite{7549086} fails to resist
impersonation attacks and man-in-the-middle attacks, as pointed out
in \cite{2017An}. Then an ECC based self-certified key distribution
scheme was proposed by Abbasinezhad-Mood \textit{et al}.
\cite{8294238} to realize higher security. But their scheme requires
the use of a tamper-proof security module. Later, Mahmood \textit{et
al}. \cite{2017Ane} presented an ECC-based authentication scheme for
the smart grid. However, their scheme fails to provide user
anonymity since the user identity is transmitted directly over the
public channel without protection. Recently, Kumar \textit{et al}.
\cite{8413131} employed ECC, symmetric encryption and hash function
to construct an authentication and key agreement scheme. Although
ECC based scheme reduces the computational costs, the EC point
multiplication operations involved in these schemes are still
time-consuming operations for smart grid environments.

Chebyshev polynomials as a lightweight operation are also adopted in
the design of authentication schemes. Chaotic maps based
authentication schemes have been applied to a variety of
environments, such as smart grids \cite{9096520}, multi-server
environment \cite{7589015,8879509}, isolated smart meters
\cite{8293795} and point-of-care systems \cite{2018Privacy0}.
Recently, Abbasinezhad-Mood \textit{et al}. \cite{0Provably}
proposed a privacy preserving authentication scheme for vehicle to
grid connections by using Chebyshev chaotic maps. Their scheme
reduces the computational cost in theory. However, they only execute
each cryptography operation independently on the devices and then
calculate a theoretical time as the execution time of their scheme.
Obviously, it is not appropriate to take the theoretical execution
time as the actual execution time of the authentication scheme.

\section{Preliminaries}
In this section, we review the basic concepts of Chebyshev chaotic and the corresponding difficult problems associated with it.

\textit{Definitionl} 1 (Chebyshev Chaotic Map): Let \textit{n} be an integer and \textit{x}$\in$[-1, 1], the Chebyshev polynomial is defined as (1) or (2) \cite{292003Public,30963463,312014An,327258361}.

\begin{flushright}
		$T_{n}(x)=cos(n\;cos^{-1}(x))\qquad\qquad\qquad(1) $ 
\end{flushright}

$T_{n}(x)\!=\!2xT_{n-1}(x)\!-\!T_{n-2}(x);\!n\!\ge\!2,\!T_{0}(x)\!=\!1,\!T_{1}(x)\!=\!x(2)$

\textit{Definitionl} 2 (Semigroup Property): One of the most important property of Chebyshev polynomial is the semigroup property, which is shown as

\begin{flushright}
	$T_{u}(T_{v}(x))=T_{uv}(x)=T_{v}(T_{u}(x))\qquad\qquad(3)$
\end{flushright}

Zhang \cite{332008Cryptanalysis} demonstrate that the semigroup property of Chebyshev polynomials also holds, when Chebyshev polynomial domain is defined on intervals $(-\infty,+\infty)$. The enhanced Chebyshev polynomial is defined as (4), where \textit{p} is a large prime number

$T_{n}(x)\!\!=\!\!(2xT_{n-1}(x)\!\!-\!\!T_{n-2}(x))mod\,p,\!n\!\ge\!2,\!x\!\!\in\!\!(-\!\infty,+\!\infty)(4)$

\textit{Definitionl} 3 (Chaotic Map-Based Discrete Logarithm Problem (CMBDLP))\cite{327258361,342016Password,35Islam2014Provably}: Given \textit{x} and \textit{y}, it is almost impossible to find the integer \textit{v}, such that $T_{v}(x) =y$. The probability that an adversary \textit{A} can solve the CMBDLP is defined as $Adv_{A}^{CMBDLP}(p)=Pr[A(x,y)=v:v\in Z_{p}^{*},y=T_{v}(x)\,mod\,p ] $.

\textit{Definitionl} 4 (CMBDLP Assumption) \cite{327258361,342016Password,35Islam2014Provably}: For any probabilistic polynomial time-bounded adversary \textit{A}, $Adv_{A}^{CMBDLP}(p)$ is negligible, that is, $Adv_{A}^{CMBDLP}(p)< \varepsilon $.

\textit{Definitionl} 5 (Chaotic Map-Based Diffie-Hellman Problem (CMBDHP))\cite{327258361,342016Password,35Islam2014Provably}: Given \textit{x}, $T_{u}\!(x)$ and $T_{v}\!(x)$, it is almost impossible to find $T_{uv}(x)$. The probability that a polynomial time-bounded adversary \textit{A} can solve the CMBDHP is defined as $Adv_{A}^{CMBDHP}\!(p)\! \!\!=\!\!Pr[A(x,T_{u}(x)\,mod\,p,T_{v}(x)\,mod\,p\!\!=\!\!T_{uv}(x)\,mod\,p\!:\!u,v\!\in\! Z_{p}^{*} ] $.

\textit{Definitionl} 6 (CMBDHP Assumption) \cite{327258361,342016Password,35Islam2014Provably}: For any probabilistic polynomial time-bounded adversary \textit{A}, $Adv_{A}^{CMBDHP}(p)$ is negligible, that is, $Adv_{A}^{CMBDHP}(p)< \varepsilon $.

\section{Chebyshev polynomial algorithm}
In this section, we describe the proposed Chebyshev polynomial algorithm in detail. In order to satisfy the security requirements and reduce the time complexity of Chebyshev polynomial algorithm, we adopt a binary exponentiation algorithm based on square matrix to compute Chebyshev polynomials. Furthermore, we employ the following matrices instead of recursive relationships to define Chebyshev polynomials.
$$\begin{gathered}
\begin{bmatrix}T_{n+1}(x)\\T_{n}(x)\end{bmatrix} =
\begin{bmatrix} 2x&-1\\1&0\end{bmatrix}\!\!\begin{bmatrix}T_{n}(x)\\T_{n-1}(x)\end{bmatrix}mod\,q
\end{gathered}$$
$$\begin{gathered}
\begin{bmatrix}T_{n+1}(x)\\T_{n}(x)\end{bmatrix}\!\!=\!\!
\begin{bmatrix} 2x&-1\\1&0\end{bmatrix}^{2}\!
\begin{bmatrix}T_{n}(x)\\T_{n-1}(x)\end{bmatrix}\!\!\!\Rightarrow \!\!\!
\begin{bmatrix} 2x&-1\\1&0\end{bmatrix}^{n}\!
\begin{bmatrix}T_{1}(x)\\T_{0}(x)\end{bmatrix}mod\,q
\end{gathered}$$
\begin{figure}[!ht]
    \vspace{-0.4cm}
    \centering
    \setlength{\abovecaptionskip}{-0.1cm}
    \includegraphics{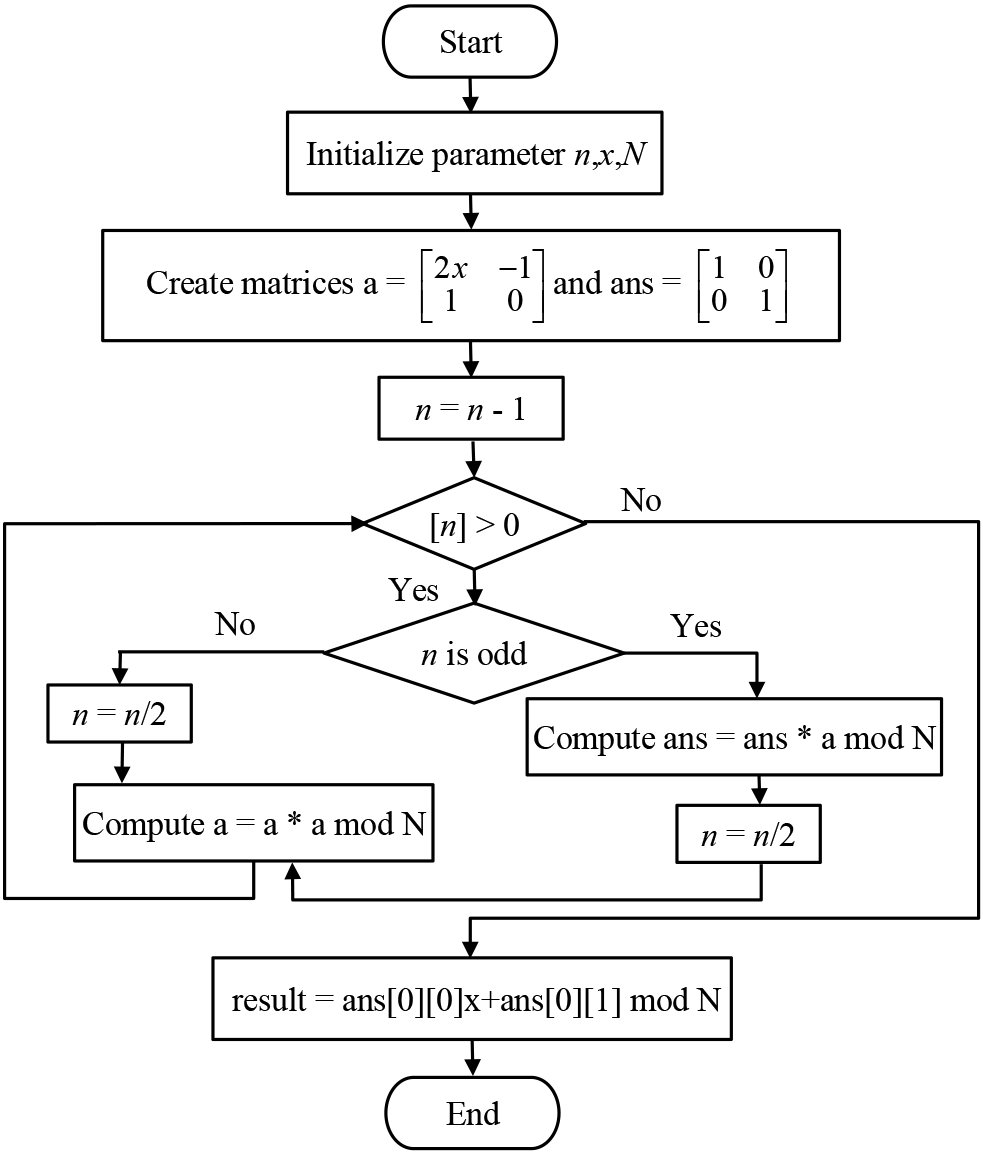}
    \caption{Flow chart of Chebyshev polynomial algorithm.}
    \label{fig2}
\end{figure}

The \textit{n}th power modulus \textit{q} of the matrix $\begin{bmatrix} 2x&-1\\1&0\end{bmatrix}$ can be solved by the binary power algorithm in polynomial time. The detailed steps of the proposed algorithm to compute $T_{n}(x)$ mod \textit{N}  are shown in Figure 2, where [\textit{n}] represents the integer part of \textit{n}.

To meet the security requirements, in our proposed Chebyshev polynomial algorithm, \textit{n} is a large prime number, and modulus \textit{N} is a strong prime number satisfying \textit{N}-1=2$p_1$ and \textit{N}+1=2$p_2$. Therefore, our proposed algorithm is safe according to \cite{2011Security12}. Furthermore, we have implemented the Chebyshev polynomial operations on Intel Intel Celeron CPU G3900T to investigate the practicability of our proposed algorithm. Table I illustrates that the proposed algorithm is an efficient Chebyshev polynomial algorithm. Furthermore, from Table I we can conclude that the execution time of Chebyshev polynomial increases as the number of bits of parameter n increases. The above conclusion also shows that the calculation method of execution time adopted in \cite{0Provably} is inappropriate.
\vspace{-0.4cm}
\begin{table}[!h]
    \setlength{\abovecaptionskip}{0cm}
    \caption{Execution Times of Chebyshev Polynomial}
    \centering
    \label{table_1}
    \begin{tabular}{c c}
        \hline\hline \\[-2.3mm]
        The bits of parameter \textit{n} & Execution time \\
        \midrule
        128bits                & 0.187385ms     \\
        160bits                 & 0.235239ms     \\
        256bits                & 0.381572ms     \\
        512bits                 & 0.754054ms     \\
        \hline\hline \\[-3.8mm]
    \end{tabular}

\end{table}

\section{Our Proposed Scheme}
Based on the proposed Chebyshev polynomial algorithm, we construct a lightweight authentication scheme that aims to reduce computational costs with high security. There are four phases in the proposed scheme, system setup phase, registration phase, login phase and authentication phase. In the system setup phase, the trusted authority (\textit{TA}) generates system parameters and publishes public information. During the registration phase, \textit{TA} completes registration of the electric vehicle ($EV_i$) with a smartcard and the aggregator ($AGT_j$) in a secure channel. Then in the authentication phase, the EV and AGT authenticate each other and establish session keys for future communications. Some notations used throughout the rest paper are described in Table II. And the procedures of the proposed scheme are presented in detail as follows.

\begin{table}[!t]
    \setlength{\abovecaptionskip}{0.cm}
    \setlength{\belowcaptionskip}{-0.9cm}
    \renewcommand{\arraystretch}{1.3}
    \caption{Execution Times of Chebyshev Polynomial}
    \centering
    \small
    \label{table_2}
    \resizebox{\columnwidth}{!}{
        \begin{tabular}{l l}
            \hline\hline \\[-3.8mm]
            \multicolumn{1}{c}{Notation} & \multicolumn{1}{c}{Definition} \\ \hline
            $ID_i$                 & Identity of the $i^th$ electric vehicle     \\
            $ID_j$                 & Identity of the $j^th$ aggregator     \\
            $x$                 & The seed of Chebyshev polynomial     \\
            $p$                 & Large prime number    \\
            $h(\cdot )$            &    Secure hash function    \\
            $r,r_i,r_j,r_u,r_s,r_1,r_2$      &    high-entropy random numbers    \\
            $k,pub_{TA}$           &   Private and public key of \textit{TA}    \\
            $k_i,k_j$           &    Private key of $EV_i$   and $AGT_j$   \\
            $SK_{ij},SK_{ji}$           &    Session key of $EV_i$   and $AGT_j$   \\
            $||$           &   The concatenation operation    \\
            $\oplus$           &    The exclusive-or operation    \\
            \hline\hline
        \end{tabular}
    }
\vspace{-0.4cm}
\end{table}
\vspace{-0.2cm}
\subsection{System Setup Phase}
In the system setup phase, the trusted authority (\textit{TA}) selects system parameters and generates its key pairs. Meanwhile, \textit{TA} registers each aggregator $AGT_j$ before their deployment in the network. The steps of this phase are given below.

\textbf{\textit{Step} 1}: The trusted authority \textit{TA} first chooses a large prime number \textit{p}, and then selects a high entropy random integer  $x\in Z_{n} ^{*} $  as the seed of Chebyshev polynomial.

\textbf{\textit{Step} 2}: The trusted authority \textit{TA} selects a high entropy random integer $ k\in Z_{n} ^{*} $ as its private key, and computes its corresponding public key $pub_{TA}$ as in (\ref{eq1}). Next, the \textit{TA} generates its identity $ID_{TA}$ and calculates a pseudo identity for itself using the value $ID_{TA}$  and its private key \textit{k} as (\ref{eq2}).
\vspace{-0.2cm}
\begin{equation}
pub_{TA}=T_{k}(x)\;mod\;p
\label{eq1}
\end{equation}
\vspace{-0.6cm}
\begin{equation}
RID_{TA}=h(ID_{TA}||k)
\label{eq2}
\end{equation}

\textbf{\textit{Step} 3}: The trusted authority \textit{TA} chooses a collision-resistant hash functions $h():{\{0,1}\}^* \longrightarrow{}{\{0,1}\}^l$. Then it publishes the system parameters {$p$,$x$, $pub_{TA}$, $h()$} and keeps its privacy key \textit{k} secretly.
\vspace{-0.2cm}
\subsection{Registration Phase}
When an electric vehicle $EV_i$ wants to access the aggregator $AGT_j$, it needs to perform the following registration process. In this phase, a smartcard is issued for each electric vehicle owner and the communication channels are supposed to be secure.

\textbf{\textit{Step} 1}: Firstly, the aggregator $AGT_j$ generates its identity $ID_j$, and selects a high entropy random integer $r_j\in Z_{n} ^{*}$. Then it calculates its pseudo identity $RID_j$ as in (\ref{eq3}) and sends it to the trusted authority \textit{TA} in a secure channel. After receiving the message from the aggregator $AGT_j$, the \textit{TA} chooses a high entropy random integer $r_s\in Z_{n} ^{*}$ and computes $Q_j$ as in (\ref{eq4}) using this integer, the received value $RID_j$, its public key $pub_{TA}$ and private key \textit{k}. Then the \textit{TA} further adopts the computed value $Q_j$, random integer $r_s$ and its private key \textit{k} to calculate the signature $s_j$ as (\ref{eq5}). After that, it sends the message \{$RID_{TA}$, $Q_j$, $s_j$\} to the aggregator $AGT_j$ through a secure channel. Subsequently, the aggregator $AGT_j$ computes its private key $k_j$ as (13) using the signature $s_j$ received from the \textit{TA}. Finally, the aggregator $AGT_j$ stores the information \{$RID_{TA}$, $RID_j$, $Q_j$, $k_j$\} in its memory secretly. When this step is finished, the registration process of the aggregator $AGT_j$ on the \textit{TA} is  completed.
\vspace{-0.2cm}
\begin{equation}
RID_j=h(ID_j\oplus r_j)
\label{eq3}
\end{equation}
\vspace{-0.6cm}
\begin{equation}
Q_j=T_{RID_j}T_{h(r_s\oplus k)}(pub_{TA})
\label{eq4}
\end{equation}
\vspace{-0.6cm}
\begin{equation}
s_j=h(RID_j||Q_j)h(r_s\oplus k)k
\label{eq5}
\end{equation}
\vspace{-0.6cm}
\begin{equation}
k_j=h(ID_j\oplus r_j)s_j
\label{eq6}
\end{equation}

\textbf{\textit{Step} 2}: The electric vehicle owner (EVO) freely chooses his/her identity $ID_i$ and password $PW_i$. Then, it selects a high entropy random integer $r_i\in Z_{n} ^{*}$ and calculates its pseudo identity $RID_i$ as in (\ref{eq7}). It also adopts its identity $ID_i$ and password $PW_i$ to compute $RPW_i$ as in (\ref{eq8}). After that, EVO sends message \{$ID_i$, $RID_i$, $RPW_i$\} to the trusted authority \textit{TA} through a secure channel. After receiving the message, the trusted authority \textit{TA} chooses a high entropy random integer $r_u\in Z_{n} ^{*}$ and then uses this integer, EVO’s pseudo identity $RID_i$, public key $pub_{TA}$ and private key \textit{k} to obtain $Q_i$ as in (\ref{eq9}). Then, the \textit{TA} further calculates the signature $s_i$ as (\ref{eq10}) via the computed value $Q_i$, pseudo identity $RID_i$, random integer $r_u$ and private key \textit{k}. Next, the \textit{TA} adopts its private key \textit{k} and the computed value $RPW_i$ to generate \textit{Y} as in (\ref{eq11}). Then \textit{TA} sends \{$ID_i$, $RPW_i$\} to the relevant $AGT_{j}$. After receiving the message, the $AGT_{j}$ computes $ A_{i}$ as (\ref{eq12}), \textit{Z} as (\ref{eq13}) and $M_{i}$ as (\ref{eq14}). Afterwards, $AGT_{j}$ sends \{$Z$, $M_i$, $RID_j$, $Q_j$\} to the \textit{TA}. After that, \textit{TA} writes \{$Z$, $M_i$, $RID_j$, $Q_j$, $RID_{TA}$, \textit{Y}, $s_i$\} into the smartcard and delivers the smartcard to the EVO in a secure way. Upon receiving the smartcard, EVO adopts \textit{Y} stored in the smartcard and its identity $ID_i$ to compute \textit{I} as (\ref{eq15}). And then it computes the private key $k_i$ as (\ref{eq16}) using its privacy information \{$ID_i$, $PW_i$, $r_i$\} and the signature $s_i$ of the \textit{TA}. Then, the EVO stores the information \{$RID_i$, $k_i$\} and replaces the \textit{Y} with \textit{I} in the memory of his/her smartcard. Finally, the memory of the smartcard contains \{$RID_{TA}$, \textit{I}, $k_i$, $Z$, $M_i$, $RID_j$, $Q_j$\}. After this step, the EV/EVO finishes the registration at the \textit{TA}.
\vspace{-0.2cm}
\begin{equation}
RID_i=h(r_i\oplus ID_i)
\label{eq7}
\end{equation}
\vspace{-0.6cm}
\begin{equation}
RPW_i=h(ID_i\oplus PW_i)
\label{eq8}
\end{equation}
\vspace{-0.6cm}
\begin{equation}
Q_i=T_{RID_i}T_{h(r_u\oplus k)}(pub_{TA})
\label{eq9}
\end{equation}
\vspace{-0.6cm}
\begin{equation}
s_i=h(RID_i||Q_i)h(r_u\oplus k)k
\label{eq10}
\end{equation}
\vspace{-0.6cm}
\begin{equation}
Y=T_{RPW_i}T_{k}(x)
\label{eq11}
\end{equation}
\vspace{-0.6cm}
\begin{equation}
A_{i} =h(ID_{i} ||k_{j} )
\label{eq12}
\end{equation}
\vspace{-0.6cm}
\begin{equation}
Z=RPW_{i} \oplus A_{i} 
\label{eq13}
\end{equation}
\vspace{-0.6cm}
\begin{equation}
M_{i} =ID_{i} \oplus h(k_{j})
\label{eq14}
\end{equation}
\vspace{-0.6cm}
\begin{equation}
I=T_{ID_i}(Y)
\label{eq15}
\end{equation}
\vspace{-0.6cm}
\begin{equation}
k_i=h(r_i\oplus ID_i||PW_i)s_i
\label{eq16}
\end{equation}
\vspace{-0.6cm}

\subsection{Login Phase}
In this phase, the registered EVO makes a login request to the aggregator $AGT_j$.

\textbf{\textit{Step} 1}: The EVO first inserts the smartcard in the $EV_i$ and inputs his/her identity $ID_i$ and password $PW_i$.

\textbf{\textit{Step} 2}: Then the electric vehicle $EV_i$ computes $RPW_i$ as (\ref{eq17}) using the inputted identity $ID_i$ and password $PW_i$ and computes $I_0$ as (\ref{eq18}) to check whether the value of $I_0$ is equal to the value of \textit{I}. If true, the electric vehicle $EV_i$ chooses a high entropy random integer $r_1\in Z_{n} ^{*}$ and calculates $C_1$ as (\ref{eq19}). Then $EV_i$ computes $A_i$ as (\ref{eq20}) and $C_2$ as (\ref{eq21}). Finally, the electric vehicle $EV_i$ sends the login request \{$C_1$, $C_2$, $M_i$\} to the corresponding aggregator $AGT_j$ via a public channel.
\vspace{-0.2cm}
\begin{equation}
RPW_i=h(ID_i\oplus PW_i)
\label{eq17}
\end{equation}
\vspace{-0.6cm}
\begin{equation}
I_0=T_{ID_i}(T_{RPW_i}(pub_{TA}))
\label{eq18}
\end{equation}
\vspace{-0.6cm}
\begin{equation}
C_1=T_{r_1}(x)
\label{eq19}
\end{equation}
\vspace{-0.6cm}
\begin{equation}
A_{i}=Z \oplus RPW_{i}
\label{eq20}
\end{equation}
\vspace{-0.6cm}
\begin{equation}
C_{2} =h(ID_{i} ||RID_{j} ||A_{i} ||C_{1})
\label{eq21}
\end{equation}
\vspace{-0.2cm}
\subsection{Authentication and key agreement Phase}
After receiving a login request from electric vehicle $EV_i$, some messages need to transmit between the electric vehicle $EV_i$ and the accessed aggregator $AGT_j$ to achieve the mutual authentication and key agreement. In this phase, the $EV_i$ and the $AGT_j$ perform the following steps:

\textbf{\textit{Step} 1}: When the aggregator $AGT_j$ receives the login request \{$C_1$, $C_2$, $M_i$\}, it first gets the identity of $EV_i$ as (\ref{eq22}) using its secret key $k_j$ and the received $M_i$. Next, it further computes $A_{i} ^{'}$  as (\ref{eq23}), $C_{2} ^{'}$ as (\ref{eq24}) and checks whether the equation $C_{2} ^{'}=C_2$ holds. If true, it chooses a high entropy random integer $r_2\in Z_{n} ^{*}$ and calculates $C_3$ as (\ref{eq25}). And the aggregator $AGT_j$ further computes \textit{X} as (\ref{eq26}). Finally, it generates an authentication message $Auth_s$ as (\ref{eq27}) and sends message \{$C_3$, $Auth_s$\} to the $EV_i$ via a public channel.
\vspace{-0.2cm}
\begin{equation}
ID_{i} =M_{i} \oplus h(k_{j})
\label{eq22}
\end{equation}
\vspace{-0.6cm}
\begin{equation}
A_{i} ^{'} =h(ID_{i} ||k_{j})
\label{eq23}
\end{equation}
\vspace{-0.6cm}
\begin{equation}
C_{2}^{'} =h(ID_{i} ||RID_{j} ||A_{i}^{'} ||C_{1} )
\label{eq24}
\end{equation}
\vspace{-0.6cm}
\begin{equation}
C_3=T_{r_2}(Q_j)
\label{eq25}
\end{equation}
\vspace{-0.6cm}
\begin{equation}
X=T_{r_2k_j}(C_1)=T_{r_1r_2k_j}(x)
\label{eq26}
\end{equation}
\vspace{-0.6cm}
\begin{equation}
Auth_s=h(X||A_{i}^{'}||ID_i||RID_j||C_3)
\label{eq27}
\end{equation}

\textbf{\textit{Step} 2}: After receiving the responding message \{$C_3$, $Auth_s$\} from the aggregator $AGT_j$, the electric vehicle $EV_i$ uses the \{$RID_j$, $Q_j$\} stored in the smart card and the received message $C_3$ to obtain ${X}^{'}$ as (\ref{eq28}). Next, the electric vehicle $EV_i$ computes ${Auth_s}^{'}$ as (\ref{eq29}) and compares it with the received authentication message $Auth_s$. If they are equivalent, it computes $C_4$ as (\ref{eq30}). After that, the electric vehicle $EV_i$ uses the computed value ${X}^{'}$ and $C_4$ to generate its authentication message $Auth_u$ as (\ref{eq31}). Finally, the electric vehicle $EV_i$ computes the session key $SK_{ij}$ as (\ref{eq32}) and sends the message \{$C_4$, $Auth_u$\} to the aggregator $AGT_j$.
\vspace{-0.2cm}
\begin{equation}
X^{'}=T_{r_1}(T_{h(RID_j||Q_j)}(C_3))=T_{r_1h(RID_j||Q_j)r_2}(Q_j)
\label{eq28}
\end{equation}
\vspace{-0.6cm}
\begin{equation}
{Auth_s}^{'}=h(X^{'}||A_{i}||ID_i||RID_j||C_3)
\label{eq29}
\end{equation}
\vspace{-0.6cm}
\begin{equation}
C_4=h(A_{i}||{X}^{'})
\label{eq30}
\end{equation}
\vspace{-0.6cm}
\begin{equation}
Auth_u=h({X}^{'}||C_4)
\label{eq31}
\end{equation}
\vspace{-0.6cm}
\begin{equation}
SK_{ij}=h(RID_{TA}||{X}^{'}||A_i)
\label{eq32}
\end{equation}
\vspace{-0.2cm}

\textbf{\textit{Step} 3}: Upon receiving the message from the electric vehicle $EV_i$, the aggregator $AGT_j$ first computes ${Auth_u}^{'}$ as (\ref{eq33}). And then it checks whether the equation ${Auth_u}^{'}$=$Auth_u$ holds. If true, it obtains the shared session key $SK_{ji}$(=$SK_{ij}$) as (\ref{eq34}).
\vspace{-0.2cm}
\begin{equation}
Auth_u^{'}=h(X||C_4)=h(T_{r_2k_j}(C_1)||C_4)
\label{eq33}
\end{equation}
\vspace{-0.6cm}
\begin{equation}
SK_{ji}=h(RID_{TA}||X||A_i^{'})
\label{eq34}
\end{equation}

Finally, the electric vehicle $EV_i$ and the aggregator $AGT_j$ achieve mutual authentication and key negotiation. The login and authentication also are shown in Figure 3.

Suppose that the $EV_i$ and the $AGT_j$ are legal. The $SK_{ij}$ is the session key generated by the $EV_i$ and the $SK_{ji}$ is the session key computed by the $AGT_j$. Now we prove that the equation $SK_{ij}$=$SK_{ij}$ is held in our proposed scheme.

\textbf{\textit{Proof}}:
$ SK_{ij}=h(RID_{TA}||{X}^{'}||A_i) $
\\$=h(RID_{TA}||T_{r_1}(T_{h(RID_j||Q_j)}(C_3))||h(ID_{i} ||k_{j})$
\\$=h(RID_{TA}||T_{r_1}(T_{h(RID_j||Q_j)}(T_{r_2}(Q_j)))||h(ID_{i} ||k_{j})$
\\$=h(RID_{TA}||T_{r_1r_2h(RID_j||Q_j)}T_{RID_j}T_{(r_s\oplus k)}\!(pub_{TA})||A_i^{'})$
\\$=h(RID_{TA}||T_{r_1r_2h(RID_j||Q_j)h(ID_j\oplus r_j)h(r_s\oplus k)k}(x)||A_i^{'})$
\\$=h(RID_{TA}||T_{r_1r_2h(ID_j\oplus r_j)s_j}(x)||A_i^{'})$
\\$=h(RID_{TA}||T_{r_1r_2k_j}(x)||A_i^{'})$
\\$=h(RID_{TA}||T_{r_2k_j}(C_1)||A_i^{'})$
\\$=h(RID_{TA}||X||A_i^{'})$
\\$=SK_{ji}$

\begin{figure}[!h]
	\vspace{-0.4cm}
	\centering
	\setlength{\abovecaptionskip}{-0.1cm}
	\includegraphics{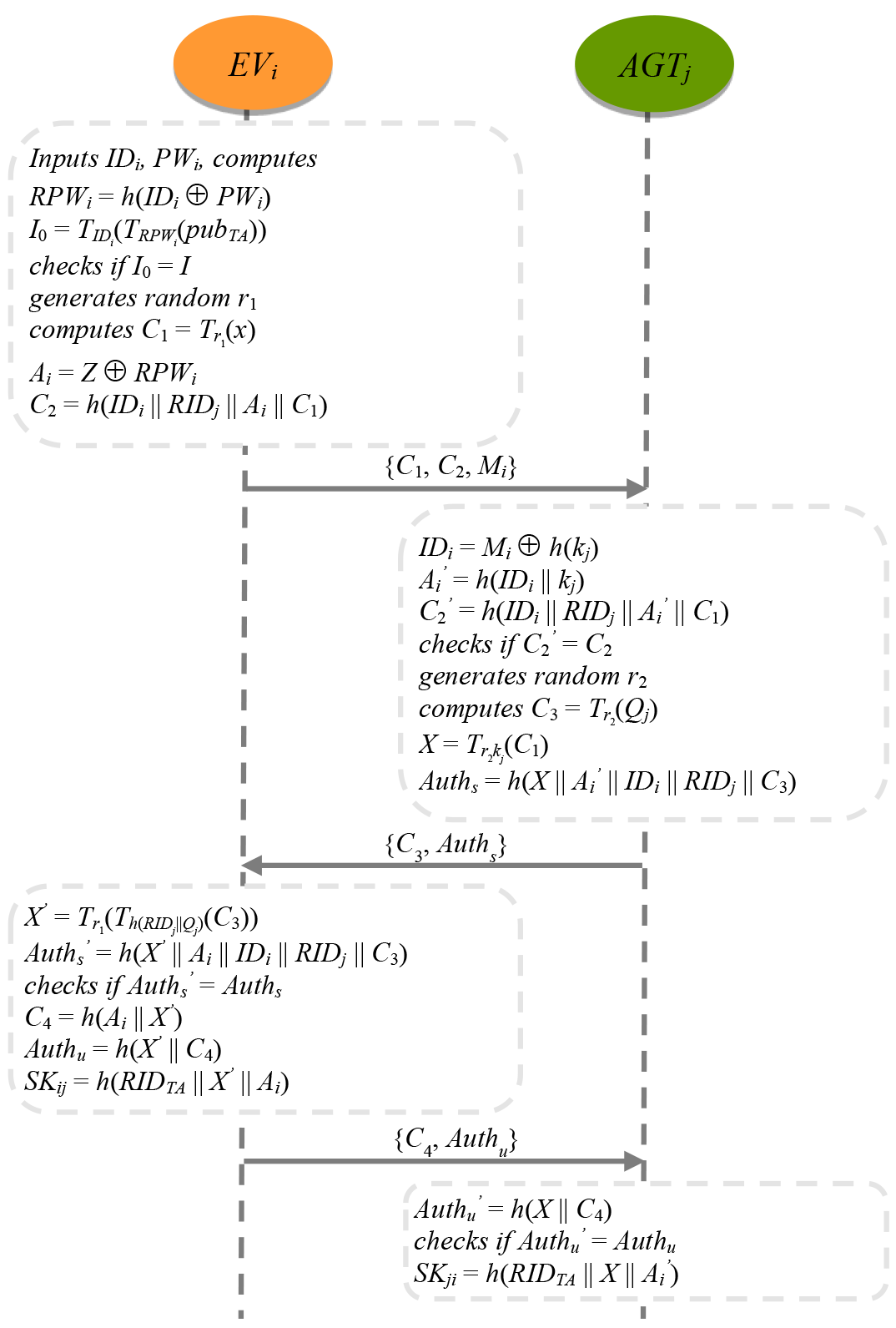}
	\caption{Illustration of login and authentication phase.}
	\label{fig7}
	\vspace{-0.2cm}
\end{figure}

\section{Security Analysis}
In this section, we adopt an automatic verifier named ProVerif to analyze the security of the proposed scheme. Moreover, we discuss the possible attacks in Section VI-B.

\subsection{Automatic Formal Verification of Security Using ProVerif}

In this section, we demonstrate the security of our proposed scheme using a widely accepted automatic protocol verifier named ProVerif \cite{36}. ProVerif can be utilized to verify the correspondence assertions, observational equivalences and reachability properties. Specifically, we can validate the resistance of cryptographic protocols against impersonation attacks, modification attacks, and replay attacks by launching injective correspondence assertion queries. Moreover, by using observational equivalence queries, some security properties such as identity guessing attacks can be verified via ProVerif. Furthermore, by making reachability queries, both the anonymity feature and the secrecy of the session key can be checked. Significantly, ProVerif can also be used to verify the perfect forward secrecy of the protocol by leaking some parameters. So, we employed ProVerif tool to implement our proposed authentication scheme and the authentication phase of the $EV_i$ and the $AGT_j$ are shown in Figure 4.
\vspace{-0.2cm}
\begin{figure}[!ht]
    \centering
    \setlength{\abovecaptionskip}{-0.2cm}
    \includegraphics{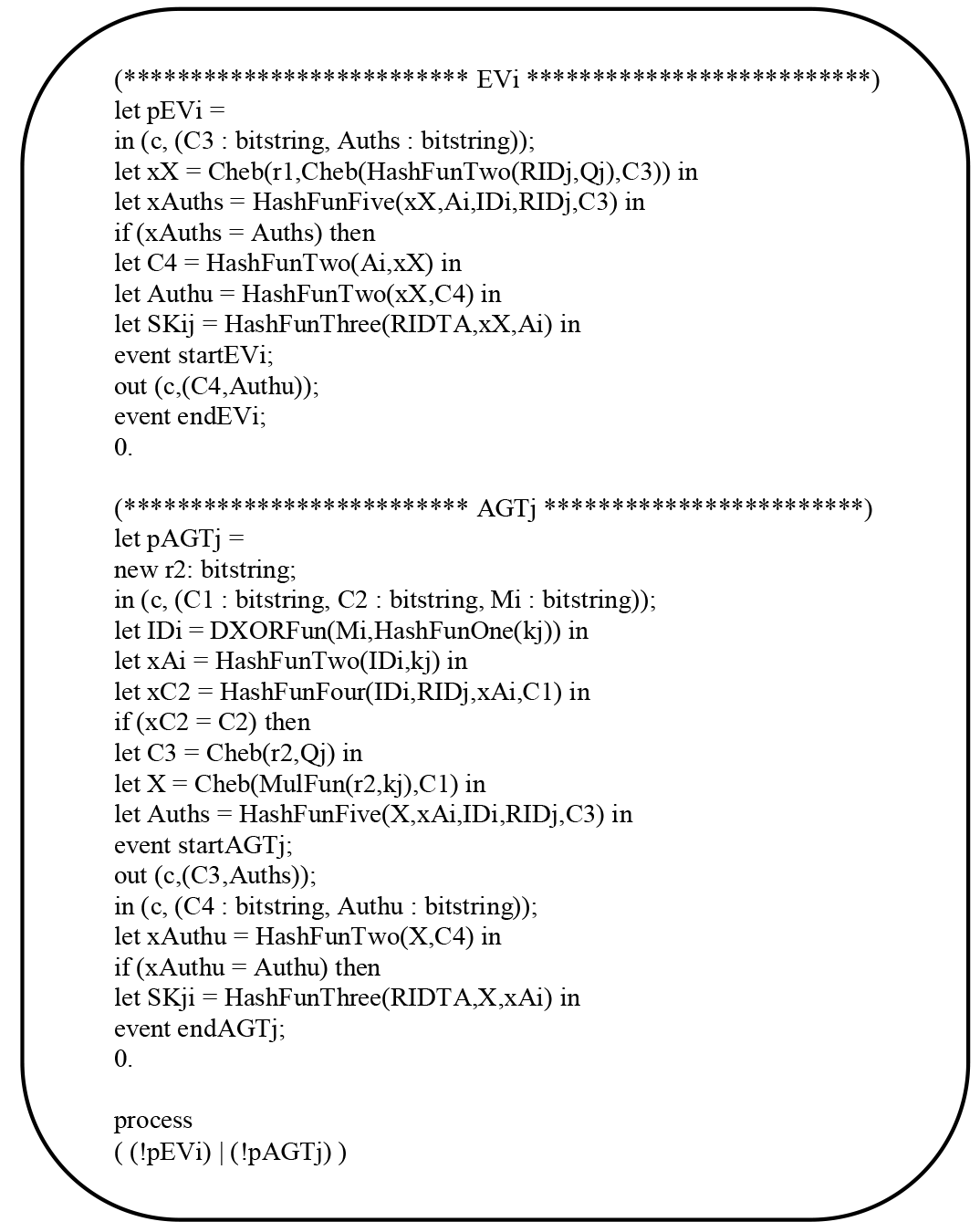}
    \caption{the authentication phase of the $EV_i$ and the $AGT_j$.}
    \label{fig3}
    \vspace{-0.2cm}
\end{figure}

Figure 5 indicates the results from the Proverif. From Fig. 5, the results (1)-(2) show that the adversary cannot obtain the private key $k_i$ of the EVi and the private key $k_j$ of the $AGT_j$. Results (3)-(4) demonstrate the secrecy of the session keys $SK_{ij}$ and $SK_{ji}$. Results (5)-(6) prove the anonymity of the $EV_i$ and the $AGT_j$. Results (7)-(8) are the results of two injective correspondence assertions which guarantees that the mutual authentication between the $EV_i$ and the $AGT_j$ is valid. In addition, injectivity allows the $EV_i$ and the $AGT_j$ to check the freshness of received messages which can resist replay attacks. Therefore, the results (1)-(8) prove that the proposed scheme provides session key security, anonymity, mutual authentication and can resist replay attacks.
\begin{figure}[!ht]
	    \vspace{-0.4cm}
    \centering
    \setlength{\abovecaptionskip}{-0.1cm}
    \includegraphics{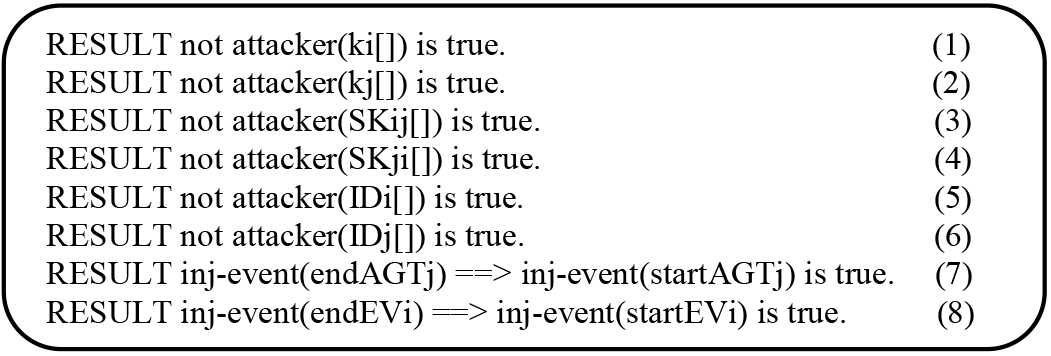}
    \caption{Results from the ProVerif (1).}
    \label{fig4}
    \vspace{-0.2cm}
\end{figure}

Moreover, we also conducted experiments to demonstrate that our proposed scheme provides perfect forward secrecy. In our experiments, the private keys of the $EV_i$ and the $AGT_j$ are transmitted over the public channel \textit{c}, which means long-term keys are leaked to the adversary. As the results (1)-(2) of  Figure 6 shows that both "not attacker(ki[])" and "not attacker(kj[])" are false, which proves that the adversary has obtained the $k_i$ and the $k_j$. However, the results of "not attacker (SKij[])" and "not attacker (SKji[])" are still true. It demonstrates that even if the $k_i$ and the $k_j$ are leaked, the session key $SK_{ij}$($SK_{ji}$) cannot be compromised. Therefore, the proposed scheme provides perfect forward secrecy.
\begin{figure}[!ht]
    \vspace{-0.4cm}
    \centering
    \setlength{\abovecaptionskip}{-0.1cm}
    \includegraphics{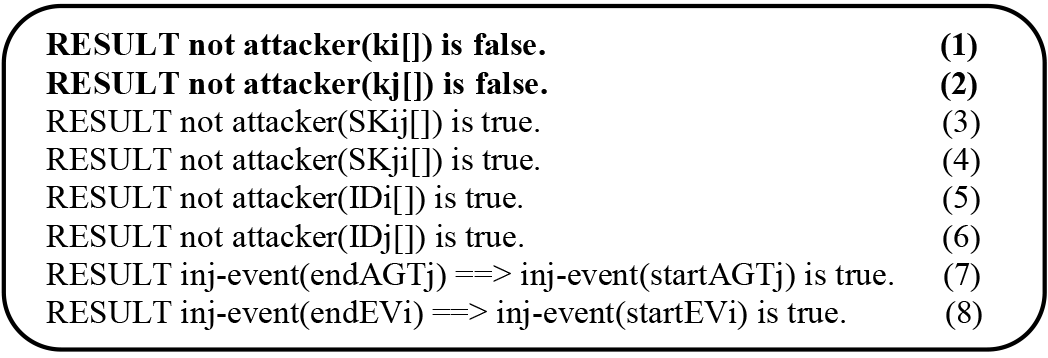}
    \caption{Results from the ProVerif(2).}
    \label{fig5}
    \vspace{-0.25cm}
\end{figure}
    \vspace{-0.5cm}
\subsection{Discussion on Possible Attacks}
In this subsection, we discuss the security of our proposed scheme by analyzing some possible attacks. In Section VI-A, we have adopted ProVerif to demonstrate that our scheme provides perfect forward secrecy, session key security, anonymity and mutual authentication and can resist replay attacks. So, in this subsection, we focus on some other attacks that we have not discussed in detail such as man-in-the-middle attacks, smartcard stolen attacks, etc.

\textbf{\textit{Man-in-the-middle attacks}}: In our scheme, the electric vehicle $EV_i$ and the aggregator $AGT_j$ can share a session key $SK_{ij}$($SK_{ji}$) only after they authenticate each other. In Section VI-A we have demonstrated that the proposed scheme achieves mutual authentication via ProVerif, so the adversary \textit{A} cannot construct independent connections with either the $AGT_j$, or the $EV_i$ . Therefore, the adversary \textit{A} cannot perform the man-in-the-middle attacks successfully.

\textbf{\textit{Modification attacks}}: Suppose that an adversary \textit{A} modifies the message sending by the $AGT_j$ with fraud \{$C_{3}^{*}$, $Auth_{s}^{*}$\} and delivers it to the electric vehicle $EV_i$ to impersonate the aggregator $AGT_j$. However, without the knowledge of the $EV_i$’s identity $ID_i$ and the $AGT_j$’s secret key $k_j$, the adversary \textit{A} cannot construct a valid $A_{i}^{'}$ to pass the verification of $EV_i$. Then the $EV_i$ will find these attacks by checking whether the equation $Auth_{s}^{'}$=$Auth_{s}$ holds.

On the other hand, assume that the adversary \textit{A} modifies the message \{$C_{1}^{*}$, $C_{2}^{*}$, $M_{i}^{*}$\} and sends it to the $AGT_j$. Similarly, if the adversary \textit{A} wants to pass the $AGT_j$’s verification, he/she needs to construct a valid $C_{2}$=$h(ID_{i} ||RID_{j} ||A_{i} ||C_{1})$. However, the adversary \textit{A} can’t generate a proper $A_{i}$ to pass the equation verification of $C_{2}^{*}$=?$C_{2}$. Therefore, our proposed scheme can resist modification attacks.

\textbf{\textit{Privileged-insider attacks}}: Assume that a privileged-insider user of the \textit{TA} is an adversary \textit{A}, and he/she has the registration message \{$RID_i$, $RPW_i$, $ID_i$\} of the $EV_i$. In the proposed scheme, the EVO’s $PW_i$ is protected by a secure hash function. So, the \textit{TA} cannot obtain the EVO’s real identity $ID_i$ and password $PW_i$, and thus cannot figure out the $I_0=T_{ID_i}(T_{h(ID_i\oplus PW_i)}(pub_{TA}))$ further. Therefore, the proposed scheme can resist privileged-insider attacks.

\textbf{\textit{Insider impersonation attacks}}: Assume a legal electric vehicle $EV_a$ becomes a malicious adversary and try to impersonate another legal $EV_i$. However, without knowing the $EV_i$’s $ID_i$, $PW_i$ and $Z$, the $EV_a$ cannot figure out the $EV_i$’s private information $A_i$ to pass the verification of the $AGT_j$. So, malicious electric vehicle $EV_a$ can’t impersonate a legal electric vehicle $EV_i$ to communicate with the $AGT_j$. On the other hand, when a registered aggregator $AGT_b$ becomes a malicious attacker and try to impersonate another legitimate aggregator $AGT_j$, he/she needs to obtain $AGT_j$’s private key $k_j$. However, without the knowledge of $AGT_j$’s $ID_j$, $r_j$ and $s_j$, the $AGT_b$ cannot calculate $k_j$ correctly. Therefore, our scheme can resist insider impersona-tion attacks.

\textbf{\textit{Offline password guessing attacks with/without smart-cards}}: Assume that an adversary \textit{A} obtains all the messages transmitting between the $EV_i$ and the $AGT_j$ and tries to launch an offline dictionary attack to get $EV_i$’s password. To obtain the $PW_i$, the adversary \textit{A} first needs to extract $A_i$ from $C_2$. Even if the adversary \textit{A} gets the $k_i$, he/she still cannot obtain PWi without the knowledge of $EV_i$’s $ID_i$, $r_i$ and $s_i$. Therefore, the adversary \textit{A} cannot launch offline dictionary attacks without smartcards successfully.

Suppose that an adversary compromises all the private information \{$RID_{TA}$, \textit{I}, $k_i$, $Z$, $M_i$, $RID_j$, $Q_j$\} stored in the smartcard of the $EV_i$ and performs offline dictionary attacks with smartcards. Compared with offline dictionary attacks without smartcards, the additional information known by the adversary \textit{A} in this attack is the information \{$RID_{TA}$, \textit{I}, $k_i$, $Z$, $M_i$, $RID_j$, $Q_j$\} stored in the smartcard. According to the above discussion, the adversary \textit{A} cannot obtain the $EV_i$’s $PW_i$ by using $k_i$. Furthermore, when the adversary \textit{A} tries to extract $PW_i$ from $I$=$T_{ID_i}(T_{h(ID_i\oplus PW_i)k}(x))$, he/she will face the CMBDLP. Even if the adversary \textit{A} solves the CMBDLP, without knowing the $EV_i$ ’s $ID_i$ and the \textit{TA}’s private key $k_i$, he/she still cannot guess $PW_i$ correctly. Thus, the proposed scheme can resist offline password guessing attacks with/without smartcards.
    \vspace{-0.2cm}
\section{Performance Analysis}
In this section, we will compare the security features, com-putational costs and communication costs of our proposed scheme with other related schemes \cite{0Provably,2017Secure,7549086,2017Ane,8413131}.
\vspace{-0.4cm}
\subsection{Comparison of Security Features}
The security features of our proposed scheme and the other five related schemes \cite{0Provably,2017Secure,7549086,2017Ane,8413131} are compared in Table III. As shown in Table III, Mahmood \textit{et al}.’s scheme \cite{2017Ane} fails to provide user anonymity. Odelu \textit{et al}.’s scheme \cite{7549086} is vulnerable to impersonation attacks and man-in-the-middle attacks. Although Wazid \textit{et al}.’s scheme \cite{2017Secure} and Kumar \textit{et al}.’s scheme \cite{8413131} are successful against common attacks, they involve time-consuming operations. Moreover, the related schemes \cite{0Provably,2017Secure,7549086,2017Ane,8413131} don’t provide automatic formal verification of security. According to Table III, our proposed scheme can resist various attacks and provide more security features in comparison with the other five related schemes \cite{0Provably,2017Secure,7549086,2017Ane,8413131}.
\begin{table}[!h]
    \vspace{-0.6cm}
        \setlength{\abovecaptionskip}{0.cm}
        \caption{SECURITY FEATURES COMPARISONS OF RELATED SCHEMES}
    \centering
    \label{table_3}
    \resizebox{\columnwidth}{!}{
    	\begin{threeparttable}
    \begin{tabular}{lcccccc}
        \hline\hline\\[-2.5mm]
        \multicolumn{1}{c}{Security attributes}                                                 & \cite{0Provably} & \cite{2017Secure} & \cite{7549086} & \cite{2017Ane} & \cite{8413131} & Ours \\ \midrule
        Replay attacks resistance                                                               & $\surd $       & $\surd $        & $\surd $        & $\surd $        & $\surd $       & $\surd $    \\
        \begin{tabular}[c]{@{}l@{}}Man-in-the-middle attacks \\ resistance\end{tabular}         & $\surd $       &$\surd $        & $\times$         & $\surd $        & $\surd $       & $\surd $    \\
        Modification attacks resistance                                                         & $\surd $      & $\surd $       & $\surd $        & $\surd $      & $\surd $        & $\surd $   \\
        \begin{tabular}[c]{@{}l@{}}Privileged-Insider attacks \\ resistance\end{tabular}        & $\surd $       & $\surd $       & $\times$        & $\surd $       & -        & $\surd $     \\
        \begin{tabular}[c]{@{}l@{}}Insider impersonation attacks \\ resistance\end{tabular}     & $\surd $      & $\surd $        & $\times$        & $\surd $        & $\surd $       & $\surd $    \\
        \begin{tabular}[c]{@{}l@{}}Offline password guessing \\ attacks resistance\end{tabular} & $\surd $       & $\surd $        & -        & -        & -        & $\surd $    \\
        Perfect forward secrecy                                                                 & $\surd $      & $\surd $        & $\surd $       & $\surd $        & $\surd $       & $\surd $   \\
        Session key security                                                                    & $\surd $      & $\surd $        & $\surd $        & $\surd $       & $\surd $        & $\surd $    \\
        Anonymity                                                                               & $\surd $      & $\surd $        & $\surd $        & $\times$        & $\surd $        & $\surd $    \\
        Low computational cost                                                                  & $\surd $       & $\times$        & $\times$        & $\times$        & $\times$        & $\surd $    \\
        Low communication cost                                                                  & $\surd $       & $\surd $        & $\times$        & $\times$        & $\times$        & $\surd $   \\
        Formal security analysis/proof                                                          & $\surd $       & $\surd $        & $\surd $        & $\surd $       & $\surd $        & $\surd $    \\
        
        \begin{tabular}[c]{@{}l@{}}Automatic formal verification \\ of security\end{tabular}    &  $\times$      & $\times$        & $\times$        & $\times$        & $\times$        & $\surd $      \\ \hline\hline
    \end{tabular}
\begin{tablenotes}
	\footnotesize
	\item $\surd $:Prevents the attack or provides the security property; 
	×: Does not prevent the attack or does not provide the property; 
	$-$: Does not discuss the security property; 
\end{tablenotes}
\end{threeparttable}
}
\vspace{-0.4cm}
\end{table}
\vspace{-0.3cm}
\subsection{Computational Cost}
In this subsection, we compare the computational costs of our proposed scheme and other five related schemes. In our experiments, we adopt OpenSSL library \cite{37}, GMP library \cite{38} and PBC Library \cite{39} to simulate these schemes on two Ubuntu 16.04 virtual machines with an Intel(R) Pentium(R) CPU G850 2.90 GHz processor, 4GB of RAM. The simulation results are shown in Table IV. The notation $T_b$, $T_e$, $T_s$, $T_h$, $T_m$, $T_a$ and $T_c$ denote the time for executing a bilinear pairing operation, a modular exponentiation operation, a symmetric key encryption/decryption operation, a one-way hash function operation, a scalar multiplication operation of an elliptic curve, a point addition operation of an elliptic curve and a Chebyshev polynomial operation, respectively.

As shown in Table IV, the Odelu \textit{et al}.’s scheme \cite{7549086} requires performing four point multiplication operations of elliptic curve, fourteen hash operations, five modular exponentiation operations and two bilinear pairing operations to complete the authentication. Then, the execution time is given by $4T_m +14T_h+5T_e+2T_{b}$ and the actual simulation time was 20.895ms. From Table IV, the computational costs of Odelu \textit{et al}.’s scheme are much higher than other related schemes \cite{0Provably,2017Secure,2017Ane,8413131} and our scheme. That because Odelu \textit{et al}.’s scheme \cite{7549086} involves heavyweight operations—bilinear pairing operations. In addition, from Table IV, the total execution time of Wazid \textit{et al}.’s scheme \cite{2017Secure}, Mahmood \textit{et al}.’s scheme \cite{2017Ane}, and Kumar \textit{et al}.’s scheme \cite{8413131} is 14.438ms, 15.764ms and 13.523ms respectively. Compared with Odelu \textit{et al}.’s scheme \cite{7549086}, these schemes \cite{2017Secure,2017Ane,8413131} reduce the computational costs effectively by avoiding the use of bilinear pairing operations.

Furthermore, the proposed scheme requires to perform five Chebyshev polynomial operations and seven hash operations on the EV side, and needs to execute two Chebyshev polynomial operations and five hash operations on the AGT side. Then the total execution time is given by $7T_c+12T_h$ and the actual simulation time is 2.807ms. As shown in Table IV, the computational costs of our scheme and Abbasinezhad-Mood \textit{et al}.’s scheme \cite{0Provably} are 2.807ms and 4.659ms, which are much lower than other related schemes \cite{2017Secure,7549086,2017Ane,8413131}. According to Table IV, our scheme and Abbasinezhad-Mood \textit{et al}.’s scheme \cite{0Provably} outperforms the related schemes \cite{2017Secure,7549086,2017Ane,8413131} in terms of the computational overhead. That is because efficient Chebyshev polynomial operations are adopted in our scheme and Abbasinezhad-Mood  \textit{et al}.’s scheme \cite{0Provably}.

\begin{table}[!h]
    \vspace{-0.4cm}
    \setlength{\abovecaptionskip}{0.cm}
    \renewcommand{\arraystretch}{1.3}
    \caption{COMPUTATIONAL COSTS COMPARISON}
    \centering
    \label{table_4}
    \resizebox{\columnwidth}{!}{
    \begin{tabular}{cccc}
        \hline\hline\\[-3mm]
        \multicolumn{1}{l}{} & \textit{EV}/$M\!D_i$/$U_i$/\textit{SM}                                               & \textit{AGT}/$S\!M_i$/$U_j$/\textit{NAN}                                                 & Total                                                          \\ \hline
        \cite{0Provably}              & \begin{tabular}[c]{@{}c@{}}$7T_h+4T_c $\\ $\approx$1.419ms\end{tabular} & \begin{tabular}[c]{@{}c@{}}$7T_h+4T_c+2T_s $\\ $\approx$3.240ms\end{tabular} & \begin{tabular}[c]{@{}c@{}}$14T_h+2T_c+8T_s $\\ $\approx$4.659ms\end{tabular} \\

        \cite{2017Secure}             & \begin{tabular}[c]{@{}c@{}}$4T_m+2T_a+5T_h $\\ $\approx$9.571ms\end{tabular} & \begin{tabular}[c]{@{}c@{}}$2T_m+8T_h$\\ $\approx$4.867ms\end{tabular} & \begin{tabular}[c]{@{}c@{}}$6T_m+2T_a+13T_h $\\ $\approx$14.438ms\end{tabular} \\

        \cite{7549086}            & \begin{tabular}[c]{@{}c@{}}$2T_m\!+\!7T_h\!+\!2T_e\!+\!2T_b $\\ $\approx$13.053ms\end{tabular} & \begin{tabular}[c]{@{}c@{}}$2T_m+7T_h+3T_e $\\ $\approx$7.842ms\end{tabular} & \begin{tabular}[c]{@{}c@{}}$4T_m\!+\!14T_h\!+\!5T_e\!+\!2T_b $\\ $\approx$20.895ms\end{tabular} \\

        \cite{2017Ane}            & \begin{tabular}[c]{@{}c@{}}$4T_m+3T_a+4T_h $\\ $\approx$9.349ms\end{tabular} & \begin{tabular}[c]{@{}c@{}}$3T_m+3T_a+4T_h $\\ $\approx$6.415ms\end{tabular} & \begin{tabular}[c]{@{}c@{}}$7T_m+6T_a+8T_h $\\ $\approx$15.764ms\end{tabular} \\

        \cite{8413131}             & \begin{tabular}[c]{@{}c@{}}$3T_m+2T_s+6T_h $\\ $\approx$6.647ms\end{tabular} & \begin{tabular}[c]{@{}c@{}}$3T_m+2T_s+7T_h $\\ $\approx$6.876ms\end{tabular} & \begin{tabular}[c]{@{}c@{}}$6T_m+4T_s+13T_h $\\ $\approx$13.523ms\end{tabular} \\

        Ours                 & \begin{tabular}[c]{@{}c@{}}$5T_c+7T_h $\\ $\approx$1.915ms\end{tabular} & \begin{tabular}[c]{@{}c@{}}$2T_c+5T_h $\\ $\approx$0.892ms\end{tabular} & \begin{tabular}[c]{@{}c@{}}$7T_c+12T_h $\\ $\approx$2.807ms\end{tabular} \\\hline\hline
    \end{tabular}
}
\vspace{-0.2cm}
\end{table}
As shown in Figure 7, the proposed scheme achieves the best performance, taking only 2.807ms in total. And on the AGT side, our scheme also achieves the lowest computational costs, which only takes 0.892ms. Compared with other related schemes \cite{0Provably,2017Secure,7549086,2017Ane,8413131}, our proposed scheme reduces the computational costs up to 39.7\%, 80.5\%, 86.6\%, 82.2\% and 79.2\% respectively. Therefore, the proposed authentication scheme is an energy efficient authentication scheme and is suitable for smart grid environments.

\begin{figure}[!ht]
    \vspace{-0.4cm}
    \centering
    \setlength{\abovecaptionskip}{-0.1cm}
    \includegraphics{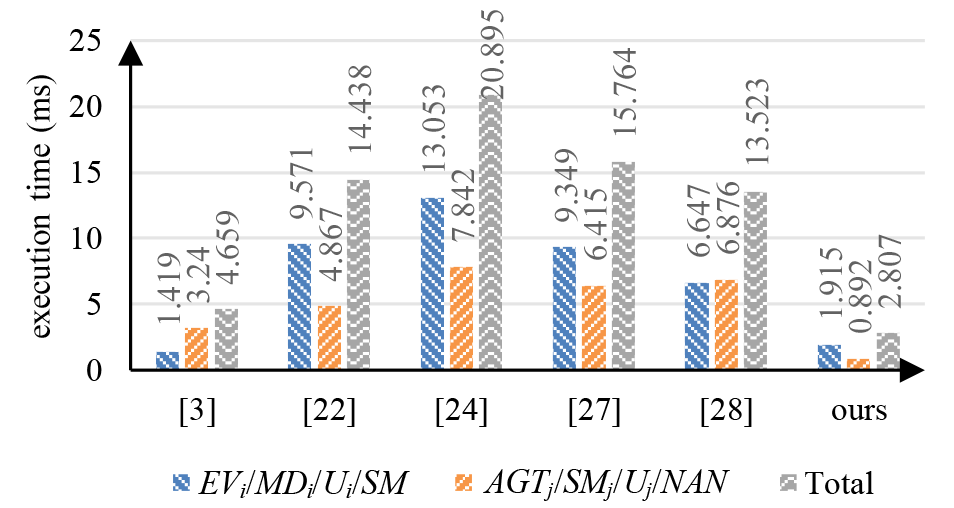}
    \caption{Comparison of execution time in the login and authentication phase.}
    \label{fig6}
    \vspace{-0.6cm}
\end{figure}

\subsection{Communication Cost}
The comparison of communication costs between our scheme and the other related schemes \cite{0Provably,2017Secure,7549086,2017Ane,8413131} is shown in Table V. In our experiment, the output of the hash function is 20 bytes (160 bits), the output of modular exponentiation operation is 32 bytes (256 bits), the output of Chebyshev Polynomial is 32 bytes (256 bits), and the output of a timestamp is 4 bytes (32 bits). In addition, for bilinear pairing, the elements in group $G_1$ and $G_2$ are 40 bytes (320 bits) and 64 bytes (512 bits).

As shown in Table V, the proposed scheme achieves the smallest communication load which is 1312 bits. And the communication costs for other related schemes \cite{0Provably,2017Secure,7549086,2017Ane,8413131} are 1376bits, 1536bits, 1920bits, 2112bits and 2240bits, respectively. Obviously, the proposed scheme reduces communication costs in comparison with the related schemes \cite{0Provably,2017Secure,7549086,2017Ane,8413131}.

\begin{table}[!h]
    \vspace{-0.4cm}
    \setlength{\abovecaptionskip}{0.cm}
    \caption{COMMUNICATION COSTS COMPARISON}
    \centering
    \label{table_5}
    \begin{tabular}{ccccccc}
        \hline\hline\\[-2mm]
        Schemes    & \cite{0Provably} & \cite{2017Secure} & \cite{7549086} & \cite{2017Ane} & \cite{8413131} & Ours \\
        \midrule
        Cost(bits) & 1376                              & 1536                               & 1920                            & 2112                            & 2240                            & 1312    \\
        \hline\hline
    \end{tabular}
    \vspace{-0.5cm}
\end{table}
\section{Conclusion}
In this paper, we proposed a new Chebyshev polynomial algorithm by
using a binary exponentiation algorithm based on square matrix. The
proposed algorithm solved the security challenge of Chebyshev
polynomial algorithm in practical application and realized secure
and efficient Chebyshev polynomial computation. Based on the
proposed algorithm, we further designed an energy-efficient
authentication and key agreement scheme for smart grid environments.
Since only lightweight Chebyshev polynomials and hash functions are
adopted during the authentication and key negotiation processes, the
proposed authentication scheme reduces the computational costs and
communication costs in comparison with the state-of-the-art schemes.
We also adopted ProVerif tool to analyze the security of the
proposed authentication scheme. The security analysis demonstrated
that our proposed authentication scheme can defend against various
attacks. Therefore, our proposed scheme is suitable for smart grid
environments due to tackling both security and performance.

\section*{Acknowledgements}
The research was financially supported by National Natural Science
Foundation of China (No.61972366, No. 61303237), the Foundation of
Key Laboratory of Network Assessment Technology, Chinese Academy of
Sciences (No. KFKT2019-003), the Foundation of  Guangxi  Key
Laboratory of Cryptography and Information Security (No.
GCIS201913), and the Foundation of Guizhou Provincial Key Laboratory
of Public Big Data (No. 2018BDKFJJ009, No. 2019BDKFJJ003, No.
2019BDKFJJ011).

\bibliographystyle{IEEEtran}
\bibliography{mybibfile}


\ifCLASSOPTIONcaptionsoff
  \newpage
\fi

\end{document}